# Could $^{18}$O-enriched water increase signal for PET proton range verification? A study using a chicken embryo model

Samuel España[1,2,3,*], Daniel Sánchez-Parcerisa[1,2,4], Paloma Bragado[2,5], Alvaro Gutierrez-Uzquiza[2,5], Almudena Porras[2,5], Carolina Gutiérrez-Neira[1,6], Andrea Espinosa[1,2], Víctor V Onecha[1,2], Paula Ibáñez[1,2], Víctor Sánchez-Tembleque[1,2], José M Udías[1,2], Luis M Fraile[1,2]

(1) Grupo de Física Nuclear & IPARCOS, Universidad Complutense de Madrid, CEI Moncloa, Madrid, Spain
(2) Instituto de Investigación Sanitaria del Hospital Clínico San Carlos (IdISSC), Ciudad Universitaria, Madrid, Spain
(3) Centro Nacional de Investigaciones Cardiovasculares (CNIC), Madrid, Spain
(4) Sedecal Molecular Imaging, Algete, Madrid, Spain
(5) Departamento de Bioquímica y Biología Molecular, Facultad de Farmacia, Universidad Complutense de Madrid, Madrid, Spain
(6) Centro de Microanálisis de Materiales, CMAM-UAM, Madrid, Spain

* Corresponding author. Email: sespana@ucm.es, phone: (+34) 91 394 47 84.

**ABSTRACT**

Purpose. Range verification of clinical protontherapy systems via positron-emission tomography (PET) is not a mature technology, suffering from two major issues: insufficient signal from low-energy protons in the Bragg peak area and biological washout of PET emitters. The use of contrast agents including $^{18}$O, $^{68}$Zn or $^{63}$Cu, isotopes with a high cross section for low-energy protons in nuclear reactions producing PET emitters, has been proposed to enhance the PET signal in the last millimeters of the proton path. However, it remains a challenge to achieve sufficient concentrations of these isotopes in the target volume.

Here we investigate the possibilities of $^{18}$O-enriched water (18-W), a potential contrast agent that could be incorporated in large proportions in live tissues by replacing regular water. We hypothesize that 18-W could also mitigate the problem of biological washout, as PET ($^{18}$F) isotopes created inside live cells would remain trapped in the form of fluoride anions (F$^-$), allowing its signal to be detected even hours after irradiation.

To test our hypothesis, we designed an experiment with two main goals: first, prove that 18-W can incorporate enough $^{18}$O into a living organism to produce a detectable signal from $^{18}$F after proton irradiation, and second, determine the amount of activity that remains trapped inside the cells.

Methods. The experiment was performed on a chicken embryo chorioallantoic membrane tumor model of head and neck cancer. Seven eggs with visible tumors were infused with 18-W and irradiated with 8-MeV protons (range in water: 0.74 mm), equivalent to clinical protons at the end of particle range. The activity produced after irradiation was detected and quantified in a small-animal PET-CT scanner, and further studied by placing *ex-vivo* tumors in a gamma radiation detector.

Results. In the acquired images, specific activity of $^{18}$F (originating from 18-W) could be detected in the tumor area of the alive chicken embryo up to 9 hours after irradiation, which confirms that low-energy protons can indeed produce a detectable PET signal if a suitable contrast agent is employed. Moreover, dynamic PET studies in two of the eggs evidenced a minimal effect of biological washout, with 68% retained specific $^{18}$F activity at 8 hours after irradiation. Furthermore, ex-vivo analysis of 4 irradiated tumors showed that up to 3% of oxygen atoms in the targets were replaced by $^{18}$O from infused 18-W, and evidenced an entrapment of 59% for specific activity of $^{18}$F after washing, supporting our hypothesis that F- ions remain trapped within the cells.

Conclusions. An infusion of 18-W can incorporate $^{18}$O in animal tissues by replacing regular water inside cells, producing a PET signal when irradiated with low-energy protons that could be used for range verification in protontherapy. $^{18}$F produced inside cells remains entrapped and suffers from minimal biological washout, allowing for a sharper localization with longer PET acquisitions. Further studies must evaluate the feasibility of this technique in dosimetric conditions closer to clinical practice, in order to define potential protocols for its use in patients.

**Keywords:** radiotherapy, proton therapy, range verification, positron emission tomography, contrast agent, $^{18}$O-enriched water, chicken embryo, head-and-neck cancer

# INTRODUCTION

The number of proton therapy facilities to treat cancer has increased considerably in the last few years (1). The main advantages of protons over other radiation therapy techniques are the large dose deposited at the end of their range, known as the Bragg peak, and the absence of dose distal to it. Therefore, the accuracy in positioning the distal edge of the proton beam is crucial for a correct dose delivery, i.e., ensuring a complete irradiation of the tumor and reducing the dose to organs at risk (2). Several techniques have been proposed for *in-vivo* proton range verification, including the use of positron emission tomography (PET) imaging to quantify and locate the $\beta^+$ isotopes produced by the proton beam through nuclear reactions inside the patient (3, 4). In addition, the detection of prompt gammas emitted from excitations of the target nuclei has been suggested (5) for proton range verification. Both techniques have been already tested in clinical studies in patients (6,7). Other studies suggested the measurement of the acoustic pressure waves generated by proton dose deposition using high frequency ultrasonic transducers (8) or the detection of radiation-induced changes in the constitution of human tissue which can be visible by MRI imaging (9). However, research related to this topic is still needed to find a precise and reliable method for *in vivo* range verification that can be implemented clinically (10).

While protons deposit their dose mainly through electromagnetic interactions, $\beta^+$ isotopes are produced via nuclear reactions. Since $\beta^+$-emitters originated by proton-induced reactions on human tissues, mainly $^{11}C$ and $^{15}O$, have a relatively high production threshold (17.9 MeV for $^{12}C(p,X)^{11}C$ and 14.3 MeV for $^{16}O(p,X)^{15}O$), they will not be formed in the proximity of the Bragg peak (11) limiting the applicability of PET imaging for proton beam range verification (12). The use of $^{13}N$ produced from $^{16}O$ has been suggested as an alternative (13) due to the lower energy threshold (5.5 MeV), but its integrated cross-section at low energies may not be high enough (14), and the 13N specific activity decreases at a fast pace, due to its shorter half-life and affected by biological washout processes.

On the other hand, isotopes such as $^{18}O$, $^{68}Zn$ or $^{63}Cu$ exhibit a high proton-induced reaction cross sections and comparatively lower reaction thresholds. Most of them are naturally occurring in trace amounts in natural elements and thus do not contribute to PET activation in human tissues. It has been suggested to use compounds with elements enriched in those isotopes as possible contrast media for proton PET range verification (15, 16) as their use may facilitate Bragg-peak localization via PET. A high concentration of the contrast in the irradiated area is required in order to provide a detectable activity that can be used for proton range verification. For that purpose, the use of $^{18}O$-enriched water (18-W) appears as an appealing approach (17, 18) as it is a suitable substance that can reach very high concentrations *in-vivo* on human tissues (19) by direct intravenous, intra-arterial or intratumoral administration to the patient. Also, this substitution does not require to modify the treatment plan computed from a CT image of the patient, as the energy deposition and biochemical effects in the target are not affected by the isotopic distribution of its constituent atoms. However, this must be verified experimentally as there might be small differences of the mean excitation energy of 18-W compared to regular water. Recently, we demonstrated the capability of 18-W for last-millimeter range verification in proton therapy by irradiating and performing a PET scan on a water phantom filled with jellified 18-W (14). The activation produced from the contrast agent showed an excellent correlation with the dose distribution up to several hours after irradiation, due to the long half-life of $^{18}F$.

In the present *in-vivo* study using 18-W as a contrast agent we hypothesize that, while cell membrane is permeable to regular water (and therefore also to 18-W), any $^{18}F$ ion produced inside the cell by activation of 18-W will remain trapped inside the cells for a relatively long time, minimizing the effects of biological washout (20). In this work, we performed an *in-vivo* study to evaluate our previous phantom results in a more realistic environment. For that purpose, a chicken embryo chorioallantoic membrane tumor model of head and neck cancer was used (21). Head and neck tumors are usually treated with radiotherapy (22), and for this type of cancers, proton therapy has demonstrated clinical efficacy and potential for toxicity reduction compared to photon therapy (23). In our experiment, the tumor was grown on top of the chicken chorioallantoic membrane (CAM) and infused with a buffer containing 18-W, prior to irradiation with protons. The produced activity was monitored after irradiation with a gamma radiation spectrometer and with a PET/CT scanner in order to evaluate the capability of this technique for enhancing *in-vivo* range verification in proton therapy.

## MATERIALS AND METHODS

**Materials.** 18-W containing 70% of $^{18}$O was purchased from a commercial vendor (Sigma-Aldrich). An $^{18}$O-enriched buffer was prepared by mixing 10X phosphate buffered saline (PBS 10X, Ionza) and 70% 18-W at a ratio of 9:1 (v/v) obtaining a physiological buffer that was used to inoculate the tumors generated in this study.

***In-vivo* chicken embryo chorioallantoic membrane tumor model.** Chick CAM assays were performed as previously described (24, 25). For these experiments we used tumorigenic HEp3 (T-HEp3) cells that were derived from a lymph-node metastasis from a head and neck squamous carcinoma patient as described in (26) and kept as a patient-derived xenograft in CAMs (27). Briefly, $2.5 \times 10^5$ T-HEp3 cells were inoculated on the chicken embryo CAM of specific pathogen-free (SPF), fertile, 10-day-old embryonated chicken eggs (Granja Santa Isabel, Spain) and tumors were grown for 7 days. On day 7, the eggs were transported to the accelerator facility, infused with W-18, and irradiated. Chicken embryo CAM experiments do not require any special additional allowance as long as the embryos are euthanized before hatching, as was done in this study.

**Irradiation facility.** All the samples were irradiated at the Centre for Micro Analysis of Materials - Universidad Autónoma de Madrid (CMAM-UAM) using the high-current electrostatic 5-MV tandetron accelerator. The proton beam was extracted from the external microbeam line (28) at an energy of 8 MeV. The optic properties of the beam in the vacuum line were controlled by two slits in the X and Y directions, manually adjustable with micrometric screws, and by two focusing quadrupole magnets (rotated 90 degrees). The beam exits the vacuum line through an 8-μm-thick Kapton window. A dedicated irradiation setup was built to accurately control the irradiation time and the beam position (29). For that purpose, a remote-control shutter was placed at the beam exit and a 3-axis robotic stage was used to hold the samples and to align them with the beam.

**Beam characterization and dosimetry.** Prior to sample irradiation, the transversal spread of the beam for a given configuration was characterized using Gafchromic EBT3 film at successive distances from the Kapton foil. Lateral spreads in the X and Y directions, namely $\sigma_x$ and $\sigma_y$, were measured using a single-gaussian model for the spot. Longitudinal variation of the spreads was fit to a second-degree polynomial for each dimension. The parameters of these fits, along with the particle stopping power in water (30) were introduced in an analytical dose-calculation engine, based on a simplified version of FoCa (31). This code (29) was used to produce suitable treatment plans for sample irradiation and to estimate the amount of absorbed dose from irradiation logs. Quality assurance of the plans was performed by comparing the expected dose with a plan-specific treatment verification film, after correcting for quenching effects (32).

**Irradiation of eggs.** Seven eggs (labeled E1-E3 for those analyzed *in-vivo* and T1-T4 for those analyzed *ex-vivo*) were irradiated with a proton beam at the CMAM facility using the same protocol as follows. Immediately prior to irradiation of each egg, 200 μL of $^{18}$O-enriched buffer were applied dropwise locally to the surface of the CAM during 5 min to facilitate its absorption within the tumor (see Figure 1). Afterwards, the egg was irradiated according to the following protocol. The eggs were situated at a distance of 5 cm measured from the shell to the beam exit window. The beam was estimated to travel through an additional gap of 2.5 cm of air (see Figure 1) between the shell window and the tumor surface. The combined energy loss of the beam in the exit window + air gap was estimated to be equal to 0.50 MeV. Range in water of 7.50-MeV protons is estimated at 0.74 mm (30), so that any activation observed under our experimental conditions would correspond to activation at the last millimeter of the path in a clinical beam.

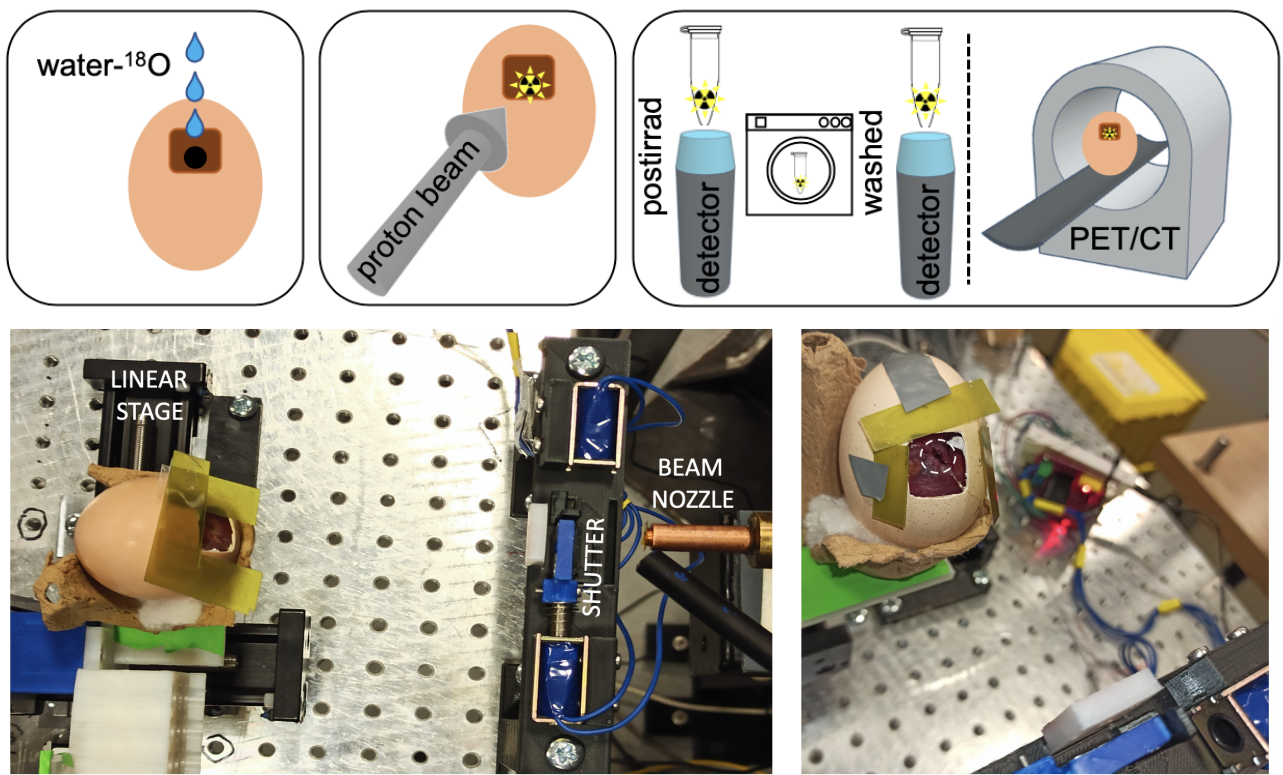

Figure 1. Top: Schematic representation of the experimental protocol followed in this study. First, the eggs were inoculated with an $^{18}O$-enriched buffer. Then, the eggs were irradiated with a proton beam of 8 MeV. Finally, 5 tumors were analyzed ex-vivo with a gamma radiation detector spectrometer and 3 eggs were analyzed on a PET/CT scanner. Bottom: Pictures of the experimental setup used for the irradiation of the eggs with the external beam line at the CMAM facility. The white circle on the right picture delineates the tumor area.

The egg was visually aligned with the beam using the robotic stage and a laser previously configured towards the beam irradiation spot (see Figure 1). In order to ensure hitting the target, we delivered 9 shots of 10 seconds duration each (90 seconds in total), with a 1.2-mm separation to each other forming a 3 × 3 grid. Spot size on the surface of the tumor was calculated at $\sigma_x$ = 0.78 mm and $\sigma_y$ = 0.89 mm. With this spot configuration, dose homogeneity in a 2-mm radius around the tumor center was better than 10%. The egg opening window was surrounded by radiochromic film to verify a correct positioning of the samples by observing the darkening of the film produced by the proton halo. The average beam intensity employed for 6 of the eggs (E1-E2, T1-T4) was 6.0 ± 0.5 nA. Tumors were irradiated with a cumulative proton fluence of about $2.5 \cdot 10^{13}$ protons/cm$^2$, depositing a total surface dose in the tumor (calculated in a thickness of 100 μm) of $2.4 \cdot 10^5$ Gy. A lower dose was delivered to one of the eggs (E3) using a much lower beam intensity of 150 pA, leading to a deposited dose of 127 Gy. These larger-than-usual doses were selected in order to ensure a clear observation of the activation effect in the samples and measure decays for a longer period, regardless of the sensitivity of the scanner, for this proof-of-principle study.

***In-vivo* analysis (PET/CT).** The activity produced on 3 of the irradiated eggs (E1 and E2 at high dose and E3 at lower dose) was monitored *in-vivo* using a SuperArgus (33) PET/CT scanner (Sedecal, Madrid, Spain). The field of view (FOV) of this scanner has a diameter of 12 cm and axial length of 10 cm, uses a pixelated phoswich detector technology combining LYSO and GSO crystals with a total thickness of 15 mm and provides a spatial resolution down to 0.8 mm FWHM. A 10-minute PET scan was performed 1 hour after irradiation for the egg irradiated at lower dose (E3) while a 9-hour PET acquisition was started 3.5-4 hours after irradiation for the eggs irradiated at higher dose (E1 and E2). Then, a CT scan was acquired. The viability of the chicken embryos was confirmed after the imaging session. The PET scan of E1 and E2 was reconstructed dividing the acquisition in 9 consecutive frames of 60 minutes, while the PET scan of E3 was reconstructed as a single frame of 10 minutes. PET images were reconstructed using the 3D-OSEM reconstruction software provided with the scanner (34) with 225 × 225 × 127 voxels and a voxel size of 0.554 mm × 0.554 mm × 0.775 mm. Tumor volumes were manually segmented on fused PET/CT images using Amide (http://amide.sourceforge.net/) and decay curves were obtained separately for E1 and E2. These curves were then fitted to an exponential following the physical decay of $^{18}$F (accounting for biological washout), as all other produced isotopes have a much shorter half-life.

***Ex-vivo* analysis (scintillator detector).** The activity produced on 4 of the irradiated eggs (T1-T4) was recorded using a gamma radiation detector composed of a CeBr$_3$ scintillation crystal with conical shape (base and top diameters of 25 and 19 mm respectively and 19 mm height) coupled to a Hamamatsu (R9779) photomultiplier tube (PMT). The detector has an energy resolution of 6% (@662 keV), and has low internal activity (35). A 3D-printed sample support was built to ensure the reproducibility of sample positioning. Within 10 min after irradiation, each tumor was excised, weighted, placed in a 0.5-mL microcentrifuge tube and shortly centrifuged. Then, the tube was placed on the detector and the gamma radiation emitted by the sample was recorded for 29 ± 5 min.

Afterwards, the tumors were washed with PBS for 5 min with gentle shaking, placed in a new tube, shortly centrifuged and placed back at the detector. Another acquisition was performed for 22 ± 13 min in order to determine the remaining activity entrapped in the tumor after washing. The gamma events recorded by the detector were processed, obtaining decay curves of the events within the 511-keV peak (±10% energy window width) and subtracting the background obtained from an acquisition with no active sample placed in the detector. The decay curves ($A_{meas}$) were fitted to a sum of exponentials (see equation 1) including the decays of $^{11}$C, $^{13}$N and $^{18}$F (see details in table I) in order to obtain the individual contribution for each isotope.

$$A_{meas}(t) = A_{N13} \sum_i exp(-\lambda_{N13} t) + A_{C11} \sum_i exp(-\lambda_{C11} t) + A_{F18} \sum_i exp(-\lambda_{F18} t) \qquad (1)$$

where $A_i$ is the initial activity (fitting parameters) and $\lambda_i$ is the decay constant for each isotope (*i*). $^{15}$O was not considered as its production cut-off energy (16.79 MeV) is well above the beam energy (8 MeV). The results obtained for each isotope were decay-corrected to the start time of the measurement performed before washing the tumor. In order to reduce the uncertainty of the fit (as the half-lives of $^{13}$N and $^{11}$C are of the same order of magnitude), a-priori information of the expected activities was employed. Specific activities of each isotope (decay-corrected to the start of the irradiation) can be expressed as:

$$A_{18F} = A_0\, N_O\, f_{18O} \int_0^{E_p} \sigma_{18O(p,n)18F}(E)\, dE\,,$$

$$A_{13N} = A_0\, N_O\, (1 - f_{18O}) \int_0^{E_p} \sigma_{16O(p,X)13N}(E)\, dE\,,$$

$$A_{11C} = A_0\, N_N \int_0^{E_p} \sigma_{14N(p,X)11C}(E)\, dE\,,$$

where $N_O$ and $N_N$ are, respectively, the number of oxygen and nitrogen atoms per unit volume, $E_p$ is the incoming proton energy, 7.50 MeV, $f_{18O}$ is the fraction of oxygen atoms of the target substituted by $^{18}$O atoms after incubation with 18-W, and $A_0$ is a constant for each irradiation encapsulating all variables which do not depend on the specific isotope, such as received dose, tumor weight and detector efficiency. From these expressions, the ratio of post-irradiation activities $A_{13N}/A_{11C}$ can be inferred from tissue atomic compositions and reaction cross sections. Assuming an atomic composition of H 9.8%, C 19.5%, N 4.8% and O 65.0 % for squamous cell carcinoma (36), and using a Monte Carlo simulation as described in previous works (14), we estimated a ratio $A_{13N}/A_{11C}$ = (1-$f_{18O}$) · 0.394. The percentage of $^{18}$F, $^{11}$C and $^{13}$N entrapped into the tumor was obtained by comparing their specific activities (decay-corrected to the irradiation time) before and after washing the tumor. Since the activity of $^{13}$N was almost negligible after wash (as post-wash measurements were performed over five $^{13}$N half-lives after irradiation ) we assumed that the activity ratio $A_{13N}/A_{11C}$ was maintained after washout or, in other words, that the washout of generated $^{11}$C and $^{13}$N was similar. This assumption has negligible impact in determining the maximum retained activity of either $^{11}$C or $^{13}$N. The percentage of incorporated $^{18}$O, or $f_{18O}$, was calculated by comparing decay-corrected specific activities of $^{18}$F and $^{11}$C before wash.

Table I. Isotopes included in the analysis of decay curves and reaction channels responsible for tumor activation (including only allowed reactions for protons of 7.5 MeV)

| Isotope | $T_{1/2}$ [min] | Reaction channel | Reaction threshold [MeV] | Integrated cross section for 7.5-MeV protons [mb·MeV] |
|---|---|---|---|---|
| $^{11}$C | 20.364 | $^{14}$N(p,X)$^{11}$C | 3.13 | 75.1 |
| $^{13}$N | 9.965 | $^{16}$O(p,X)$^{13}$N | 5.55 | 3.3 |
| $^{18}$F | 109.77 | $^{18}$O(p,n)$^{18}$F | 2.57 | 1130.7 |

## RESULTS

***In-vivo* analysis (PET/CT).** The $^{18}$F activation was monitored *in-vivo* on 3 of the irradiated eggs using PET/CT imaging. The chicken embryos remained viable during the entire experiment including irradiation and PET/CT acquisitions. Figure 2A shows an estimation of the dose distribution delivered to E1 overlaid with the CT and the PET activation images. The image is oriented in the beam's eye view. Fused CT and PET images for E1, E2 and E3 are shown on figure 2 (b-d). The PET images for E1 and E2 correspond to the first hour of acquisition and the PET image for E3 corresponds to the entire 10 min acquisition. PET images show a high activity signal on the irradiated tumors placed on the CAM, while only a minor background of diffused activity is observed in the rest of the egg or in the chicken embryo. The opening window in the eggshell can be observed on the CTs of two of the eggs.

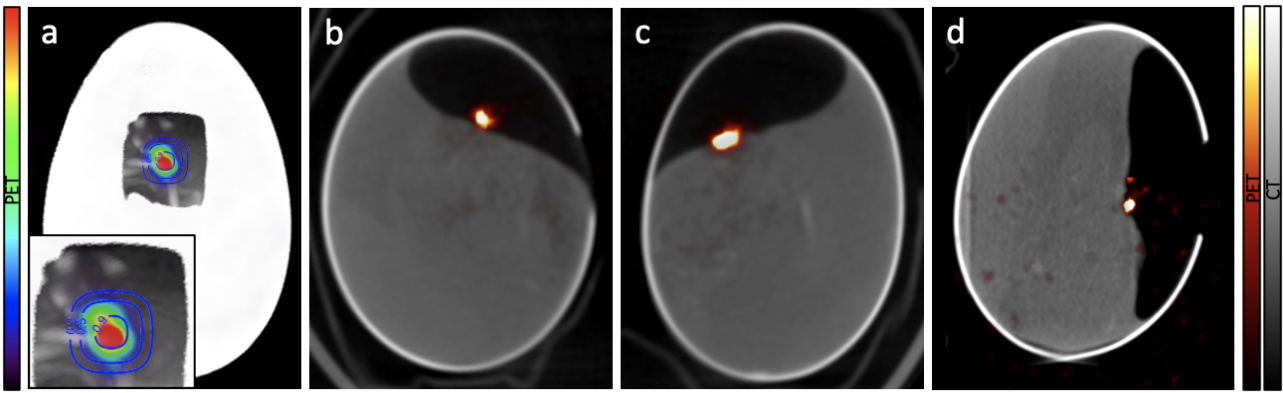

*Figure 2. (a) Overlay of an estimation of the isodose contours of the delivered dose, the CT and the PET activation for E1. A zoomed image of the irradiated area is shown in the insert. Fused PET and CT images obtained for E1 (b), E2 (c) and E3 (d).*

The decay curves obtained from the dynamic PET images for E1 and E2 are shown on Figure 3. The curves were fitted to an exponential function corresponding to the physical decay of $^{18}$F multiplied by a biological decay function (37). Combined half-lives derived from the fits were (97.6 ± 1.6) min and (101 ± 8) min for E1 and E2, yielding wash-out half-lives of (14.7 ± 2.2) and (21 ± 18) hours respectively. Therefore, an important result is that the detected activity decays mostly from the disintegration of $^{18}$F, i.e., little washout is observed for many hours: the model predicts a biological decrease in $^{18}$F activity of less than 32% at 8 hours after irradiation.

From this result, we can infer that the produced $^{18}$F ions remain largely entrapped within the tumor cells, held by the electronegative potential of the cell membrane.

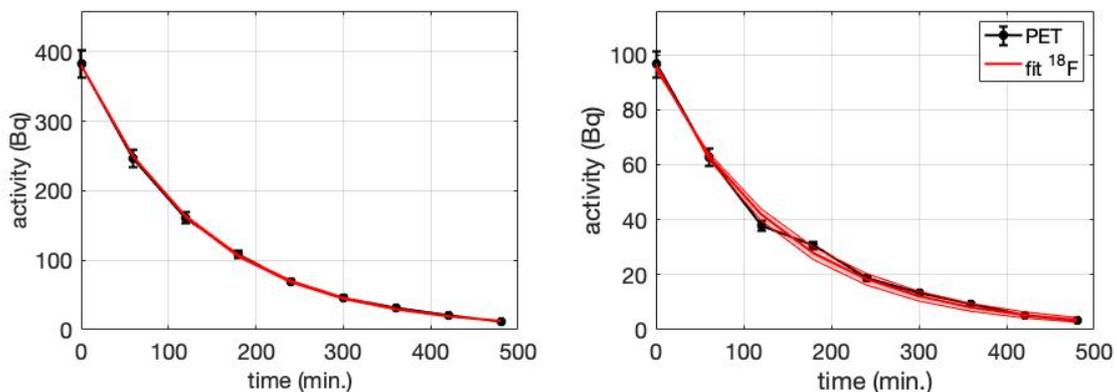

*Figure 3. Decay curves of E1 (left) and E2 (right) obtained from dynamic PET images (black) and the fits (red). The red shaded area represents 95% confidence interval of the fit.*

Figure 4 displays another activity map (with increased contrast) showing some $^{18}$F activity away from the target area. This activity corresponds to 18-W present in the extracellular space (or blood vessels) at the time of irradiation. This $^{18}$F is created outside the cell membrane and reaches other areas of the chicken embryo through diffusion or circulation. Moreover, some of this radioactive fluorine appears to cluster in the bony structures of the chicken embryo. Indeed, $^{18}$F-NaF is a well-known radiotracer used to monitor osteogenic activity for the detection of osseous metastases (38).

***Ex-vivo* analysis (CeBr$_3$ detector).** The tumors from 4 of the irradiated eggs (T1-T4) were excised after the irradiation and analyzed with the CeBr$_3$ detector before and after washing the tumor with PBS. Figure 5 shows the decay curves measured with the gamma detector for the irradiated eggs and the fits obtained with the contribution of different isotopes. Results are shown for measurements performed before and after washing the tumor with PBS. A large production of $^{18}$F can be observed before washing the tumor with contributions from $^{11}$C and $^{13}$N.

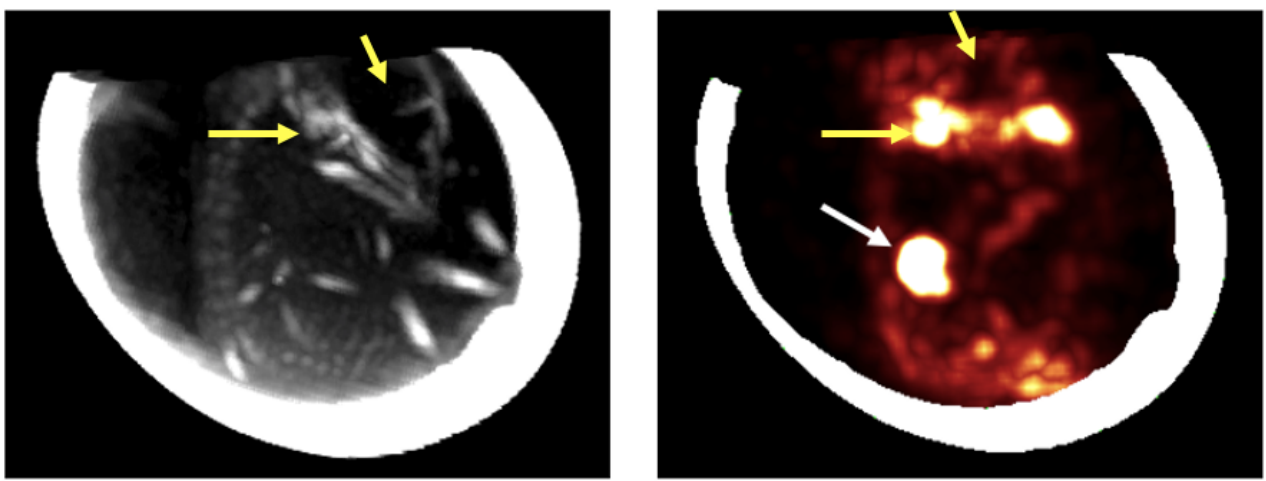

*Figure 4. CT (left) and PET (right) maximum intensity projection images of E1. The bony structure of the chicken embryo is observed on the CT and the bone uptake of the produced $^{18}$F is observed on the PET image (yellow arrows show the jaw joint with high uptake and the cranial cavity with no uptake) as well as the activity produced in the tumor (white arrow). PET image is overlaid with the eggshell. PET and CT images could not be overlaid due to motion of the chicken embryo.*

Figure 6 (top) shows absolute specific activity of the three considered isotopes at the beginning of each measurement, obtained from the fits. Differences in activity between T1-T4 tumors can be explained by tumor size, $^{18}$O captation, and time elapsed between irradiation and start of measurement. Estimated captation levels (in unwashed tumors) ranged from a minimum of 0.61% in T1 to a maximum of 3.0% in T4, expressed as the fraction of $^{16}$O atoms replaced by $^{18}$O, or $f_{18O}$.

Figure 6 (bottom) shows the percentage of retained activity of the different isotopes considered, calculated as the ratio of activities corrected by the physical decay, analyzed *ex-vivo* with the gamma detector. The weighted average shows a retention of (59±9)% for $^{18}$F, significantly higher than the estimated retention of $^{11}$C and $^{13}$N, calculated at (14±26)%. The high uncertainty in this number is due to the shorter half-lives of $^{11}$C and $^{13}$N, as most of their activity had physically decayed by the time the post-wash measurements were performed. Still, the results for the washed tumors highlight that a large portion of the produced $^{18}$F remains entrapped in the tumor while all other isotopes have been washed out or have physically decayed.

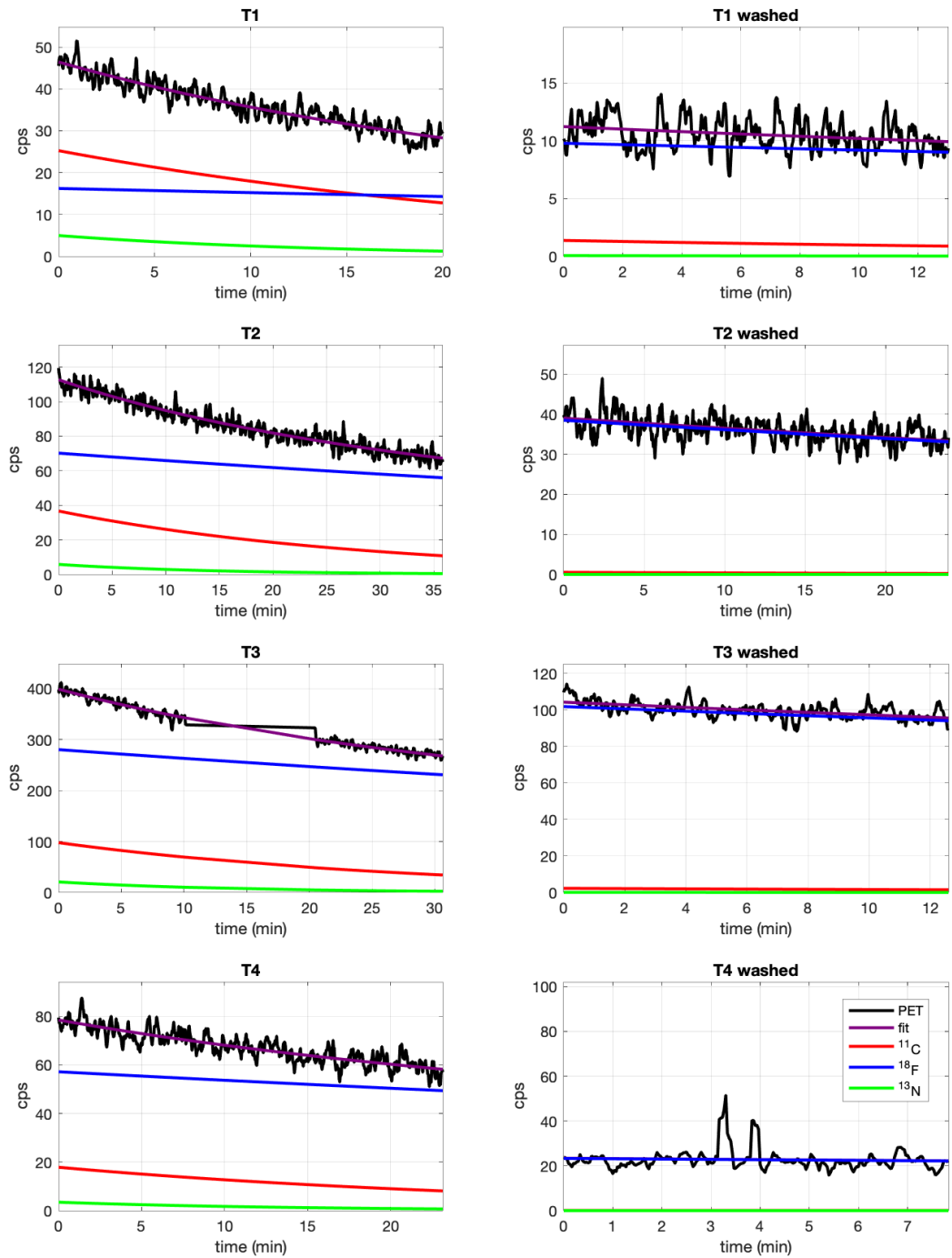

*Figure 5. Decay curves (black) of gamma events recorded with the CeBr$_3$ detector from tumors T1-T4 inoculated with an $^{18}$O-enriched buffer and irradiated with a proton beam before (left) and after (right) washing the tumor with PBS. Decay curves were fitted (purple) to a sum of exponentials to obtain the contribution from $^{11}$C (red), $^{13}$N (green) and $^{18}$F (blue). Measurement of T3 activity was stopped for 10 minutes and then resumed, hence the gap in activity shown in the plot.*

## DISCUSSION

***Analysis of the observed results.*** In this study, the proton activation in the last millimeter of the proton path was observed *in-vivo* by PET imaging. For that purpose, we inoculated 18-W into the tumor of a chicken embryo CAM tumor model of head and neck cancer prior to proton irradiation inducing the production of $^{18}$F that gets entrapped inside the tumor and can be later detected by PET imaging. Furthermore, we observed for the first time the biodistribution of *in-vivo* produced $^{18}$F radiotracer in the bony structures of the chicken embryo (see Figure 4).

The use of 18-W for proton range verification was first proposed by Cho et al (16,17,18) using implantable markers or hydrogels encapsulating 18-W and its feasibility was tested in phantoms. Instead, we proposed the direct administration to the patient of 18-W on a previous study (14). In the present work we test this approach in an animal model for the first time. Several range verification techniques have already been tested in patients (3,4,39) although they are still not used in the clinical routine.

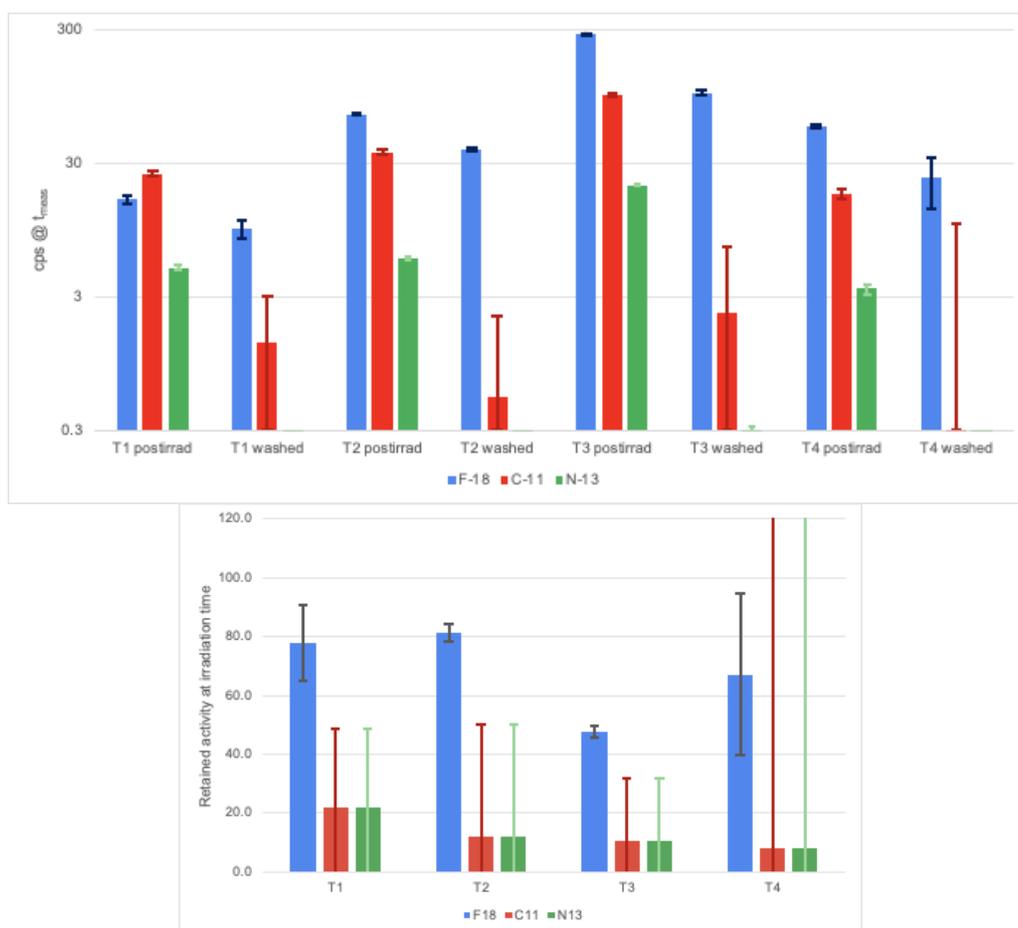

*Figure 6. (top) Measured activity in counts per second (cps) from tumors T1-T4 corresponding to $^{11}$C (red), $^{13}$N (green), and $^{18}$F (blue) at the beginning of the measurements ($t_{meas}$), performed at post-irradiation time and on washed tumors. (bottom) Percentage of retained activity on the irradiated tumors (T1-T4) obtained by comparison of the decay-corrected specific activity at the irradiation time before and after washing the tumors. Error bars represent the 1-sigma confidence interval.*

The low energy threshold in the production of $^{18}$F (2.6 MeV, proton range in water of 115 $\mu$m) induces sample activation in very close proximity to the end of the proton path, enabling more accurate range verification than other isotopes, such as $^{11}$C and $^{15}$O, which are not produced in the last few mm. The 8-MeV proton beam was used to irradiate the implanted tumor onto the CAM, previously inoculated with an $^{18}$O-enriched buffer. This led to 7.5-MeV protons entering the tumor with a range in water of 740 $\mu$m, ensuring that dose deposition and activation occurred only at the equivalent of the distal end of a clinical beam. Results shown in Figure 6 demonstrate that most of the remaining activity in the tumor a few minutes after irradiation corresponds to $^{18}$F and more than half of the produced $^{18}$F is entrapped within the tumor. Long-term entrapment of $^{18}$F was confirmed *in-vivo* by the dynamic PET images obtained for two of the irradiated eggs (E1 and E2).

Determination of retained activity was measured with reasonable accuracy (better than 10%) for $^{18}$F, but with not so good accuracy for $^{11}$C and $^{13}$N, due to the long time elapsed between irradiation and post-wash measurements. However, available data allows us to calculate an upper limit on the activity retention for $^{11}$C and $^{13}$N at 40%, proving that $^{18}$F is retained in the cells at a significantly higher quantity than other produced isotopes. Still, further studies with shorter delay times between pre-wash and post-wash measurements would be required to increase the accuracy of the entrapment measurements for short-lived isotopes.

The mechanism for the entrapment of $^{18}$F within the tumor may be explained as follows. Water is a diffusible agent that is not extracted or entrapped within the cells but rather freely diffuses across membranes (40). Once the $^{18}$F ion is produced, it no longer has the ability to cross the cell membrane due to its charge, and gets intracellularly entrapped if produced within the cytoplasm (41). Although the anionic channels in the cells are not highly specific, it is known that they show only a very small permeability to the fluoride ion (42). In addition, $^{18}$F in aqueous solution forms hydrogen bonds with the surrounding water molecules and becomes unreactive for nucleophilic substitution (43).

***Remaining challenges for clinical application.*** In order to produce enough $^{18}$F activity allowing for its visualization using PET scanners, a high concentration of 18-W within the tumor is required. In this study, the $^{18}$O-enriched buffer was directly applied on the tumor as is typically performed for drug administration in the chicken CAM model. For clinical application, different routes of administration and irradiation protocols must be explored to ensure maximal concentration of 18-W in the tumor and surrounding tissues, while using a reasonable amount of this costly contrast agent. $^{18}$O-water is regularly used all around the world as a target for production of $^{18}$F for PET imaging. In addition, doubly labeled water ($^2$H$_2$$^{18}$O) has been used for decades for measuring energy expenditure in free-living animals (44) and in humans (45). In any case, we must explore techniques that allow obtaining and using $^{18}$O-water at a reduced cost. The dose delivered to the tumor in this study was noticeably higher that in a typical clinical fraction (~2 Gy) and the sensitivity of the preclinical PET scanner is about 5 times higher than that of a clinical PET scanner. In this work, usable PET images were still obtained from 4 to 8 hours after irradiation (where initial activity had decayed by factors of 4.5 and 20, respectively, as displayed in Figure 7) and E3 was irradiated with a much lower dose than E1 and E2 and the activation was clearly visible one hour after irradiation (see Figure 2D).

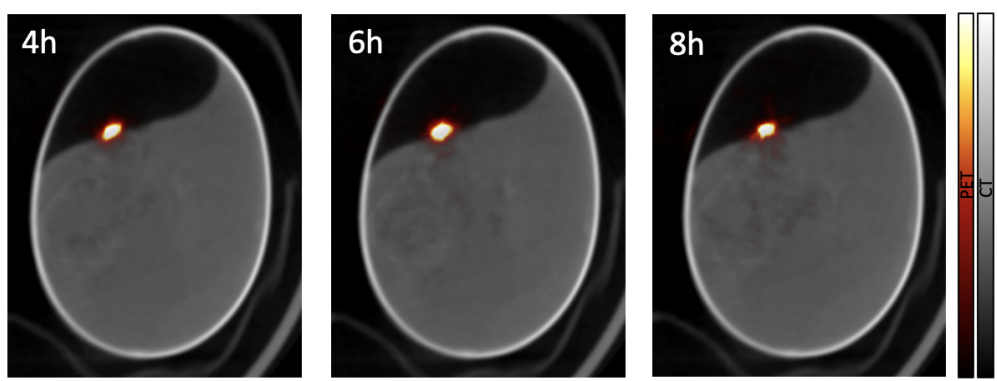

*Figure 7. Fused PET and CT images obtained for E1 acquired 4 (left), 6 (middle), and 8 (right) hours after irradiation. The color scale of the PET images was adjusted for each particular image.*

In order to estimate the clinical feasibility of the proposed technique, we performed a Monte Carlo simulation using TOPAS (46). A proton beam of 8 MeV was used to deliver a dose of 2 Gy at the Bragg Peak to a target containing 30% of 18-W achieving a $^{18}$F activity of 0.5 Bq/mm$^3$. A minimum detectable activity of 148 Bq/mm$^3$ has been reported for clinical PET scanners (47). However, the latest generation of PET scanners (48, 49) provide many technological improvements including increased sensitivity, excellent time-of-flight resolution and advanced reconstruction algorithms which may allow for the detection of lower activity concentration. Furthermore, those estimations were made for standard clinical scan duration (3-5 min) and signal to background ratio while PET scans for proton range verification can be extended for 20-30 min in a single bed position and there is no activity out of the irradiated volume. Finally, there is ample room for improvement in $^{18}$O captation (estimated at below 3% for this study), suggesting that lower doses could be explored if captation is increased. Therefore, future experiments should evaluate the feasibility of the proposed technique in an environment closer to the clinical conditions.

## CONCLUSIONS

The use of 18-W as a suitable contrast agent for *in-vivo* range verification in proton therapy has been evaluated *in-vivo* in a chicken embryo CAM tumor model of head and neck cancer. Results show $^{18}$F activation and retention within the tumor in the last millimeter of the proton range, which enables direct proton range measurement using offline PET imaging. The longer half-life of $^{18}$F makes it possible to detect it in readily available PET scanners more than 2 hours after irradiation to minimize the contribution from other isotopes with high production thresholds while a large fraction of the produced $^{18}$F remains entrapped within the tumor cells.

The observed results encourage us to proceed with further *in-vivo* experiments in larger animals and with clinically relevant proton beam energies to validate and assess the capabilities of 18-W as a suitable contrast agent for range verification in proton therapy. In order to achieve a high-enough $^{18}$O concentration in the irradiated volume, different routes of administration and irradiation protocols will be also studied.

**Acknowledgements:** This work was funded by Comunidad de Madrid under project B2017/BMD-3888 PRONTO-CM "Protontherapy and nuclear techniques for oncology." Support by the Spanish Government (MCIU/AEI, FEDER, EU) (RTI2018-098868-B-I00, RTC-2015-3772-1, SAF2016-76588-C2-1-R, RTC2019-007112-1), European Regional Funds and the European Union's Horizon 2020 research and innovation programme under the Marie Sklodowska-Curie grant agreement No 793576 (CAPPERAM) and BBVA (becas Leonardo 2018, BBM-TRA-0041) is acknowledged. This is a contribution for the Moncloa Campus of International Excellence, "Grupo de Física Nuclear-UCM", Ref. 910059. Part of the calculations of this work were performed in the "Clúster de Cálculo para Técnicas Físicas", funded in part by UCM and in part by EU Regional Funds. S. España is supported by the Comunidad de Madrid (2016-T1/TIC-1099) and A. Gutierrez-Uzquiza is supported by the Comunidad de Madrid (2017-T1/BMD-5468). The CNIC is supported by the Instituto de Salud Carlos III (ISCIII); the Ministerio de Ciencia e Innovación; and the Pro CNIC Foundation, and is a Severo Ochoa Center of Excellence (SEV-2015-0505). The authors acknowledge the support of SEDECAL staff to perform the PET and CT images in their facilities and the CMAM technical staff for their assistance during the experiments.

**Conflict of interest statement:** There are no conflicts of interest that the authors should disclose.